\documentclass[nofootinbib,showpacs,preprintnumbers,amsmath,amssymb,floatfix]
{revtex4-1}
\usepackage{epsf}

\begin{document}

%\begin{center}

\title{QCD factorization for forward hadron scattering at high energies}

\vspace*{0.3 cm}

\author{B.I.~Ermolaev}
\affiliation{Ioffe Physico-Technical Institute, 194021
 St.Petersburg, Russia}
\author{M.~Greco}
\affiliation{Department of Physics and INFN, University Roma Tre,
Rome, Italy}
\author{S.I.~Troyan}
\affiliation{St.Petersburg Institute of Nuclear Physics, 188300
Gatchina, Russia}

\begin{abstract}
We consider the QCD factorization of DIS structure functions at small $x$ and amplitudes of
$2 \to 2$ -hadronic forward scattering at high energy. We show that both collinear and $k_T$-factorization
for these processes can  be obtained approximately as reductions of a more general (totally unintegrated)
form of the factorization. The requirement of ultraviolet and infrared stability of the
factorization convolutions allows us to obtain restrictions on the fits for the parton distributions
in  $k_T$- and collinear factorization.
\end{abstract}


\maketitle

\section{Introduction}\label{intro}
QCD factorization is the fundamental concept to provide
theoretical grounds for applying Perturbative QCD to the
description of hadronic reactions. According to factorization,
any scattering amplitude $A$ in QCD can be represented as a
convolution of perturbative (E) and non-perturbative (T)
contributions. In particular, the representation of the amplitude
$A$ for the elastic Compton scattering

\begin{equation}\label{comptscatt}
\gamma (q) + h(p) \to \gamma (q) + h(p)
\end{equation}
off
a hadron target is
\begin{equation}\label{factgen}
A = \sum_r E_r \otimes T_r ,
\end{equation}
where $r$ denotes the type of
intermediate partons in the convolutions: quarks or gluons.
Similarly, the amplitudes $A_h$ of $2 \to 2$ -hadronic reactions

\begin{equation}\label{hscatt}
h_1 (p_1) + h_2 (p_2) \to h'_1 (p'_1) + h'_2 (p'_2)
\end{equation}
can be represented in a general
form through more involved
convolutions:

\begin{equation}\label{hfactgeneral}
A_h = \sum_{rr'}\widetilde{T}_r \otimes A^{(pert)}_{rr'} \otimes T_{r'}
\end{equation}
where $T$ and $\widetilde{T}$ are  parton distributions and
$A^{(pert)}_{rr'}$ is the perturbative amplitude, with the purely gluonic contribution
$A^{(pert)}_{gg}$ dominating in forward kinematics at very high energies. \\
There are two kinds of QCD factorization
in the literature: Collinear factorization\cite{colfact,efr} and $k_T$- factorization\cite{ktfact}.
For instance, the DIS structure functions $f(x,Q^2)$
are respectively represented in those factorizations as follows:

\begin{equation}
\label{colfact}
f(x,Q^2) = \int_x^1 \frac{d \beta}{\beta} f^{(pert)}(x/\beta,Q^2/\mu^2) \phi (\beta, \mu^2)
\end{equation}

and

\begin{equation}
\label{ktfact}
f(x,Q^2) = \int_x^1 \frac{d \beta}{\beta} \int
\frac{d k^2_{\perp}}{k^2_{\perp}} f^{(pert)}(x/\beta,Q^2/k^2_{\perp}) \Phi (\beta, k^2_{\perp})
\end{equation}
where $f^{(pert)}$ stands for the perturbative components of the structure functions
both in collinear and $k_T$-factorization while $\phi$ and $\Phi$ denote the involved parton
distributions.
In the present paper we
discuss only general features of these perturbative components,
so throughout the paper we will keep for them the same generic
notation $f^{(pert)}$ in both collinear and $k_T$-factorization. In contrast, we are going to
discuss details
of the involved parton distributions, so, in order to avoid any misunderstanding,
we have used in  Eqs.~(\ref{colfact},\ref{ktfact}) different notations
for the parton distributions in collinear and  $k_T$-factorization, $\phi$ and $\Phi$,
respectively.  Besides collinear and $k_T$- factorization, we will
introduce in Eq.~(\ref{fsns})
a more general factorization which we call basic factorization. This factorization
involves
new, totally unintegrated parton distributions which we will denote
$\Psi$.  The parameter $\mu$ in Eq.~(\ref{colfact}) denotes the factorization scale of
collinear factorization.
Originally, this scale parameter
was introduced a kind of border between the perturbative and non-perturbative
domains of QCD. In addition, it plays the role of the cut-off for the infrared-divergent
perturbative contributions. Also, it is often associated with the starting point of  $Q^2$-evolution.
Collinear factorization Eq.~(\ref{factgen}) treats the intermediate partons
%in the perturbative terms in the convolutions
%,\ref{factgen})
as
%considered as
nearly on-shell ones. In particular, this takes place in
DGLAP\cite{dglap}  and its generalizations (see e.g. the overview \cite{egtiree}) to the small-$x$ region.
The collinear factorization cannot be used when the perturbative contributions to
Eq.~(\ref{factgen}) are calculated using BFKL\cite{bfkl}, where the external gluons are
kept essentially off-shell. In order to embrace this case, collinear factorization was replaced by  $k_T$ -factorization.
$k_T$- and collinear factorization, being introduced for such different motivations, look unrelated to each other.
In Ref.~\cite{egtfact} we showed that both factorizations for the DIS structure functions
can be derived from a more general factorization, which we addressed as the Basic form of QCD factorization,
where the convolutions are totally unintegrated. \\
In the present paper we continue to investigate the relations
between $k_T$- and collinear factorization in more detail
than was done in Ref.~\cite{egtfact}, paying more attention to
the gauge invariance of the amplitudes involved and considering a
more complicated dependence of the parton distributions on the
transverse momenta. We first consider DIS structure functions
and then generalize the results obtained for them to the more
intricate convolutions describing factorization of the
amplitudes of $2 \to 2$-hadron scattering. Throughout the paper we
focus on the high-energy (small-$x$) domain.\\
 Our paper is
organized as follows: In Sect.~\ref{sect2} we consider the
derivation of the collinear and $k_T$-factorization for the
amplitudes of forward Compton scattering off a hadron target
and then proceed to the amplitudes of the forward hadron
scattering in Sect.~\ref{sect3}.  In Sect.~\ref{sect4} we compare
the form of collinear factorization obtained in
Sects.~\ref{sect2},~\ref{sect3} with the conventional form of this
factorization and relate these two forms. In Sect.~\ref{sect5} we
discuss the ultraviolet behavior of the totally unintegrated
parton distributions and formulate the requirements for the
factorization convolutions to be UV-stable; applying these
requirements to the standard fits in Sect.~\ref{sect6}, we
argue against the use the singular factors $x^{-a}$ in the fits. We
summarize our results in Sect.~\ref{sect7}.

\section{Factorization for forward Compton scattering}\label{sect2}

\begin{figure}[h]
\begin{center}
\begin{picture}(140,140)
\put(0,0){\epsfbox{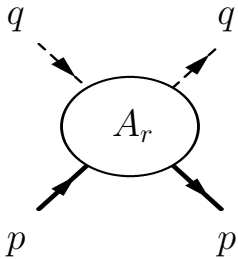} }
\end{picture}
\end{center}
\caption{\label {hfactfig1} Forward Compton scattering off a
hadron target.}
\end{figure}

The scattering amplitude $A_{\mu\nu} (p,q)$ of forward Compton
scattering off a hadron target is depicted in Fig.~\ref{hfactfig1}

%%%%%%%%%%%%%%%%%%%%%%%%%%%%%%%%%
%%%%%%%%%%%%%%%%%%%%%%%%%%%%%%%
where the blob signifies the presence of perturbative
 and non-perturbative QCD contributions. We have used the standard notations: $p$ stands for the
hadron momentum and $q~ (-q^2 = Q^2)$ is the virtual photon momentum. In what follows we omit
dependence of $A_{\mu\nu}$  on the hadron mass and spin.
Due to the Optical theorem, the imaginary part $\Im A_{\mu\nu}$  is
proportional to the hadronic tensor $W_{\mu\nu}$. Using the standard projection operators,
$A_{\mu\nu}$   can be expanded into a
set of invariant amplitudes $A_r$,
every DIS structure function $f_r$ being expressed through a certain $A_r$:

\begin{equation}\label{opt}
f_r = \frac{1}{\pi} \Im A_r.
\end{equation}

\begin{figure}[h]
\begin{center}
\begin{picture}(420,120)
\put(0,0){\epsfbox{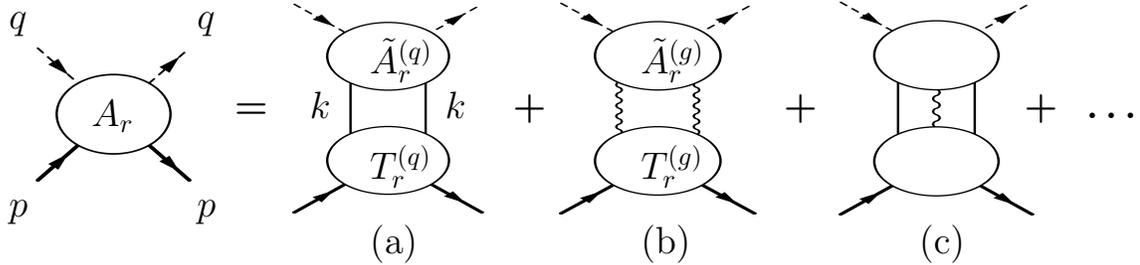} }
\end{picture}
\end{center}
\caption{\label {hfactfig2} Representation of Compton amplitude through convolutions.}
\end{figure}

The next step is to represent $A_r$ in a factorized form as an
infinite set of $t$-channel convolutions, each with a certain
number of the intermediate $t$-channel partons  (quarks and
gluons) as shown in Fig.~\ref{hfactfig2}. Each of the blobs in
Fig.~\ref{hfactfig2} can contain both perturbative and
non-perturbative contributions. In the present paper we consider
only the simplest convolutions (a) and (b) in Fig.~\ref{hfactfig2}
with two-quark and two-gluon intermediate states respectively,
denoting them by $A^{(q)}_r$ and $A^{(g)}_r$ respectively. In other
words, we assume that

\begin{equation}\label{aaqag}
A_r \approx A^{(q)}_r + A^{(g)}_r .
\end{equation}
Let us notice that our assumption is in full agreement with the conventional approach to factorization.
On the other hand, neglecting the contributions with multi-parton intermediate states could
rise questions about the gauge invariance of  $A^{(q)}_r$ and $A^{(g)}_r$. We discuss the
gauge invariance of them in the present Sect.
We will focus on discussing $A^{(q)}_r$ while
the convolution $A^{(g)}_r$ with two-gluon intermediate state can be studied quite
similarly (see Ref.~\cite{egtfact} for detail).
In the analytical form, $A^{(q)}_r$ is

\begin{equation}\label{aq}
A^{(q)}_r (p,q) = \int \frac{d^4 k}{(2\pi)^4}  \frac{\hat{k}
\tilde{A}_r (q,k)\hat{k} T_r(p,k)}{k^2 k^2}
\end{equation}
where $\tilde{A}_r~(T_r)$ denotes the upper (lowest) blob  in
Fig.~\ref{hfactfig2}(a) and $k$ stands for the intermediate quark
momentum.

%%%%%%%%%%%%%%%%%%%%%%%%%%%%%%%%%
%%%%%%%%%%%%%%%%%%%%%%%%%%%%%%%

\begin{figure}[h]
\begin{center}
\begin{picture}(280,140)
\put(0,0){\epsfbox{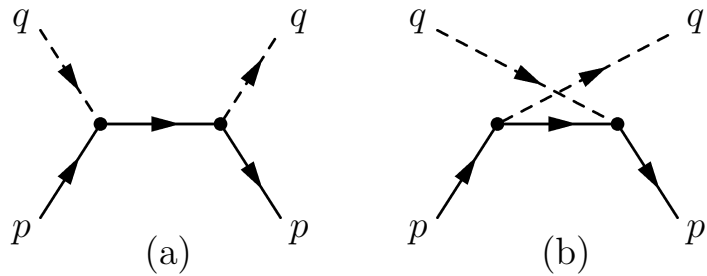} }
\end{picture}
\end{center}
\caption{\label {hfactfig3} The Born contribution to Compton
amplitude.}
\end{figure}

Each of  $\tilde{A}_r$ and $T_r$ can contain both perturbative and
non-perturbative contributions, so the factorized form of $A_r$ in
Eq.~(\ref{aq}) does not correspond to the concept of QCD
factorization where the perturbative and non-perturbative
contributions should be separated. Because of this reason we
refer to Eq.~(\ref{aq}) as the Primordial Convolution. The
integration in Eq.~(\ref{aq}) runs over the whole phase space and
includes both the ultraviolet (UV) and infrared (IR) regions. The perturbative
contributions in $\tilde{A}_r$ are IR-divergent and must be
regulated. In the first place, there is the IR-divergent power
contribution $2pk/k^2$. Such a contribution appears in one of the
amplitudes
$\tilde{A}_r$ with vacuum quantum numbers in the $t$-channel.
Throughout the paper we will refer to such an amplitude as the
singlet amplitude and denote it by $A_S$, with the upper blob
$\tilde{A}_S$ and the lower blob $T_S$, omitting the superscript
$q$. Through the Optical theorem $A_S$ is related to the singlet DIS
structure function  $F_1$. We will refer to all other
amplitude as non-singlet ones and generically denote them
$A_{NS}$, with the upper and lower blobs $\tilde{A}_{NS}$ and
$T_{NS}$ respectively. We stress that some of such
non-singlet amplitudes are related to the flavor singlet structure
functions (for example, $F_2, g_1^S$, etc). Besides the power
IR-dependent term, there are IR-divergent logarithmic
contributions $\sim \ln^n (2pk/k^2)$ and $\ln^n (Q^2/k^2)$. Such
contributions exist in the singlet and non-singlet amplitudes. It
was shown in Ref.~\cite{egtfact} that in order to regulate the IR
divergences in $A_S$ and $A_{NS}$ without explicitly cutting off the
region of small $k^2$ from the integration region in Eq.~(\ref{aq}),
the amplitudes $T_S, T_{NS}$ should obey

\begin{equation}\label{tsnsk}
T_{NS} \sim (k^2)^{\gamma},~~~T_S \sim (k^2)^{1 + \gamma}
\end{equation}
at small $k^2$, with $\gamma > 0$. It is convenient to rewrite Eq.~(\ref{aq}) in terms of
the standard Sudakov variables\cite{sud}:

\begin{equation}\label{sud}
k = -\alpha (q + xp) + \beta (p - x'q) + k_{\perp},
\end{equation}
where $x = -q^2/w, ~ x'= p^2/w$ and $w = 2pq$. Doing so, we arrive at

\begin{equation}\label{aqsud}
A^{(q)}_r (p,q) = \frac{w}{32 \pi^3}\int d \alpha d \beta d k^2_{\perp}
\frac{\hat{k} \tilde{A}_r (q,k)\hat{k} T_r(p,k)}{(w \alpha\beta + k^2_{\perp})^2},
\end{equation}
where $r = S, NS$. Now let us consider $A^{(q)}_r (p,q)$ in the
Born approximation. The Compton amplitude off the quark (the upper
blob) in the Born approximation is depicted in
Fig.~\ref{hfactfig3}.

%%%%%%%%%%%%%%%%%%%%%%%%%%%%%%%%%
%%%%%%%%%%%%%%%%%%%%%%%%%%%%%%%

The quark amplitudes $A^{(q)}_{S,NS}$ and  the gluon amplitudes
$A^{(g)}$ in the Born approximation were investigated in detail in
Ref.~\cite{egtfact}, so we omit this in the present paper. Let us
stress that the lower blobs $T_S$ and $T_{NS}$ in the Born
approximation are altogether non-perturbative. As is known, the
Born Compton amplitude is gauge-invariant when the quark is
on-shell. In Appendix A we demonstrate that gauge invariance
for those amplitude is restored in the high-energy region where $w
\gg Q^2, k^2$.

\begin{figure}[h]
\begin{center}
\begin{picture}(340,150)
\put(0,0){\epsfbox{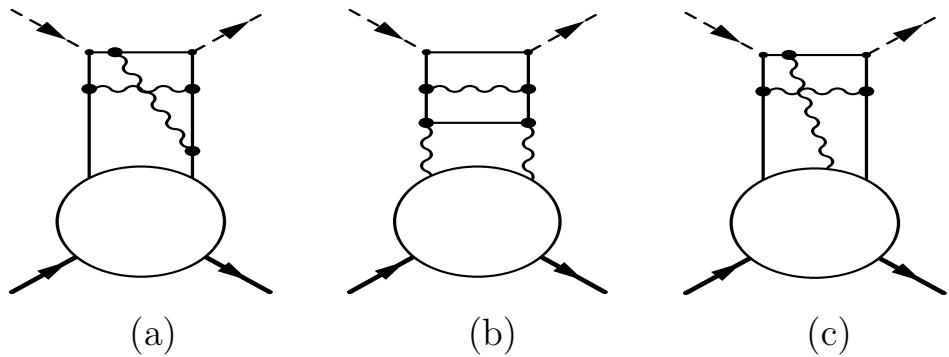} }
\end{picture}
\end{center}
\caption{\label {hfactfig4} Examples of the graphs contributing to Compton amplitude beyond the Born approximation.}
\end{figure}

Going beyond the Born approximation means adding extra gluon and
quark propagators in all possible ways to the Born graphs.
Examples of this are shown in Fig.~\ref{hfactfig4}.

%%%%%%%%%%%%%%%%%%%%%%%%%%%%%%%%%
%%%%%%%%%%%%%%%%%%%%%%%%%%%%%%%

The resulting graphs can be classified into two groups:\\
\textbf{(i)} The graphs where extra propagators do not affect the lower blob
(for instance the graphs (a) and (b) in Fig.~\ref{hfactfig4}). \\
\textbf{(ii)} The graphs where the extra propagators involve the
lower blob (graph (c) in Fig.~\ref{hfactfig4}).

Obviously, accounting for the whole set of graphs from group
\textbf{(i)} leads to the convolution depicted in
Fig.~\ref{hfactfig2}(a). whereas graphs from the group
\textbf{(ii)} form convolutions with more complicated intermediate
states. Such graphs cannot be simply neglected
%straightforwardly
because this affects the gauge invariance. However, we
demonstrate in Appendix A that gauge invariance is restored at small $x$
even when the group (\textbf{ii)} is neglected. Let us stress that we have arrived
at the convolution where the upper blob $A_r$ is altogether perturbative and at
the same time the lower blob $T_r$ is totally non-perturbative.

Applying the Optical theorem to Eq.~(\ref{aqsud}), simplifying the spinor
structure and adding the similar gluon contribution (see Ref.~\cite{egtfact} for detail)
leads to the convolution for the DIS structure functions:

\begin{equation}\label{fsns}
f_{S,NS} (x,Q^2)= \int d \alpha \frac{d \beta}{\beta}
\frac{d k^2_{\perp}}{(w \alpha\beta + k^2_{\perp})}
 f_{S,NS}^{(pert)} (x/\beta, Q^2/k^2) \Psi_{S,NS}(w\alpha, k^2)
\end{equation}
where $f_{NS}^{(pert)} ~(f_{S}^{(pert)})$ is the perturbative component of $f_{NS}~ (f_S)$
and $\Psi_{NS} = \Im T_{NS},~\Psi_{S} = \Im T_{S}$. In order to regulate the
IR-divergences in Eq.~(\ref{fsns}), these parton distributions should
decrease with $k^2$ at small $k^2$:

\begin{equation}\label{psisnsk}
\Psi_{NS} \sim (k^2)^{\gamma},~~~\Psi_S \sim (k^2)^{1 + \gamma}.
\end{equation}

Eq.~(\ref{fsns}) represents the structure functions in factorized form,
with perturbative and non-perturbative contributions being separated, so
this representation authentically corresponds to the concept of  QCD factorization.
On the other hand, this factorization involves three integrations, which distinguishes it
from both $k_T$ -factorization (which involves integration over $\beta$ and $k_{\perp}$)
and the collinear factorization (involving integration over $\beta$ only).
In Ref.~\cite{egtfact} we introduced the term "Basic form of factorization" for such
totally unintegrated convolutions
and will use it throughout the present paper. The basic factorization involves the totally
unintegrated parton distributions. In this respect we are close to Ref.~\cite{stasto} but
in contrast to that paper, we focus on small-$x$ kinematics and go beyond the Born approximation.

The basic factorization can be approximately reduced to the $k_T$ -factorization if the restriction

\begin{equation}\label{abk}
w \alpha\beta  \ll k^2_{\perp}
\end{equation}
is accepted. In this case one can perform the integration over $\alpha$  without
dealing with $f^{(pert)}$ and arrive at the $k_T$ -factorization for $f_S,~f_{NS}$:

\begin{equation}\label{fsnsktfact}
f_{S,NS} (x,Q^2) \approx \int^1_x \frac{d \beta}{\beta}
\int_0^w
\frac{d k^2_{\perp}}{k^2_{\perp}}~
 f_{S,NS}^{(pert)} (x/\beta, Q^2/k^2_{\perp})~ \Phi_{S,NS}(\beta, k^2_{\perp}),
\end{equation}
$\Phi_S$ and $\Phi_{NS}$ being the parton distributions for the $k_T$ -factorization.
They are related to the totally unintegrated parton distributions $\Psi_S,~\Psi_{NS}$:

\begin{equation}\label{phisnsktfact}
\Phi_{NS} = \int^{k^2_{\perp}/w\beta}_{s_2^{(0)}/w} d \alpha \Psi_{NS} (w\alpha, k^2),~~
%\Phi_{NS} = \int^{k^2_{\perp}/w\beta}_{k^2_{\perp}/w} d \alpha \Psi_{NS} (w\alpha, k^2),~~
%\Phi_S = \int^{k^2_{\perp}/w\beta}_{k^2_{\perp}/w} d  \alpha \Psi_S (w\alpha, k^2)
\Phi_S = \int^{k^2_{\perp}/w\beta}_{s_2^{(0)}/w} d  \alpha \Psi_S (w\alpha, k^2).
\end{equation}
The lowest limit of integrations in Eq.~(\ref{phisnsktfact}) is fixed from the following
considerations: the invariant energy $s_2 = (p -k)^2$ of $\Phi_{NS}$ and $\Phi_{S}$ must be positive, so
\begin{equation}\label{stwo}
s_2 = (p -k)^2
% w \alpha  - (w\alpha\beta + k^2_{\perp}) + p^2
\approx w\alpha - k^2_{\perp} > s_2^{(0)} > 0 .
\end{equation}

In order to regulate the IR divergences in the $k_T$-convolutions, these parton distributions should
behave  (cf. (\ref{tsnsk})) as
\begin{equation}\label{psisnskt}
\Phi_{NS} \sim (k^2_{\perp})^{\gamma},~~\Phi_{S} \sim (k^2_{\perp})^{1 + \gamma}
\end{equation}
at small $k^2_{\perp}$. Obviously, the $k_T$-factorization in Eq.~(\ref{fsnsktfact})
authentically coincides with the conventional form of
$k_T$-factorization in Eq.~(\ref{ktfact}).

\begin{figure}[h]
\begin{center}
\begin{picture}(200,200)
\put(0,0){\epsfbox{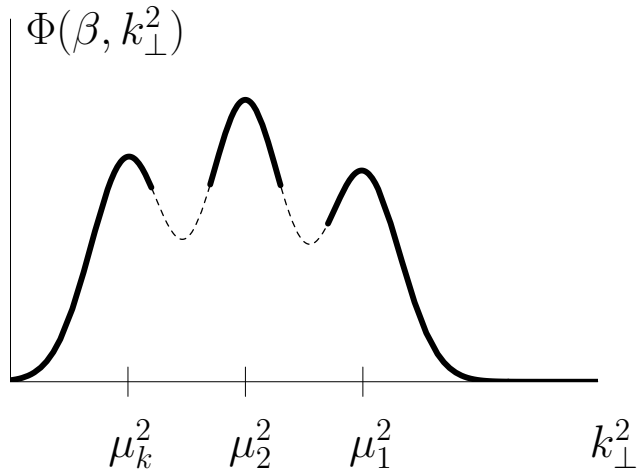} }
\end{picture}
\end{center}
\caption{\label {hfactfig5} Assumed dependence of the parton distributions on
$k^2_{\perp}$.}
\end{figure}

In its turn, $k_T$ -factorization can also be approximately
reduced to collinear factorization. In order to do it, we
assume the peaked dependence of the parton distributions on
$k^2_{\perp}$ as shown in Fig.~\ref{hfactfig5}.

%%%%%%%%%%%%%%%%%%%%%%%%%%%%%%%%%
%%%%%%%%%%%%%%%%%%%%%%%%%%%%%%%

We stress that this hypothesis can be checked by analysis of
experimental data. The peaked dependence $\Phi_{S,NS}$ on
$k^2_{\perp}$ makes it possible to perform the integrations over
$k^2_{\perp}$ in Eq.~(\ref{fsnsktfact}) without involving the
perturbative components $f^{(pert)}_{S,NS}$, arriving at the
collinear factorization

\begin{equation}\label{fsnscolfact}
f_{S,NS} (x,Q^2) \approx \sum_{k}\int^1_x \frac{d \beta}{\beta}
 f_{S,NS}^{(pert)} (x/\beta, Q^2/\mu_k^2)~ \phi_{S,NS}(\beta, \mu_k^2),
\end{equation}
with the singlet and non-singlet parton  distributions
$\phi_{NS}(\beta, \mu_k^2)$ and $\phi_{S}(\beta, \mu_k^2)$ defined
at the factorization scale $\mu_k$. They are related to the $k_T$
-parton distributions in the following way:

\begin{equation}\label{phisnscolfact}
\phi_{NS} = \int_{D_k} d k^2_{\perp} \Phi_{NS} (w\alpha, k^2_{\perp}),~~
\phi_S = \int_{D_k} d k^2_{\perp} \Phi_S (w\alpha, k^2_{\perp}).
\end{equation}
The integration regions $D_k$ in Eq.~(\ref{phisnscolfact}) are
located around the positions of the maxima of the parton
distributions at $k^2_{\perp}=\mu^2_k$ as shown in
Fig.~\ref{hfactfig5}. Obviously, Eq.~(\ref{fsnscolfact}) differs in many respects from
the conventional expression in Eq.~(\ref{colfact}) for the DIS structure functions in collinear factorization.
We will show how to bring Eq.~(\ref{fsnscolfact}) to the conventional form in Sect.~IV.

\section{Forward scattering of hadrons}\label{sect3}

\begin{figure}[h]
\begin{center}
\begin{picture}(160,120)
\put(0,0){\epsfbox{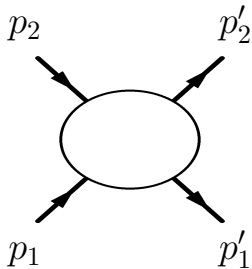} }
\end{picture}
\end{center}
\caption{\label {hfactfig6} Forward $2 \to 2$
scattering of hadrons.}
\end{figure}

Let us consider the $2 \to 2$-scattering of hadrons (see Eq.~(\ref{hscatt}))
in forward kinematics where $s = (p_1 + p_2)^2 \gg t = (p'_1 -
p_1)^2 $. This process is depicted in Fig.~\ref{hfactfig6}.

%%%%%%%%%%%%%%%%%%%%%%%%%%%%%%%%%
%%%%%%%%%%%%%%%%%%%%%%%%%%%%%%%

\begin{figure}[h]
\begin{center}
\begin{picture}(300,300)
\put(0,0){\epsfbox{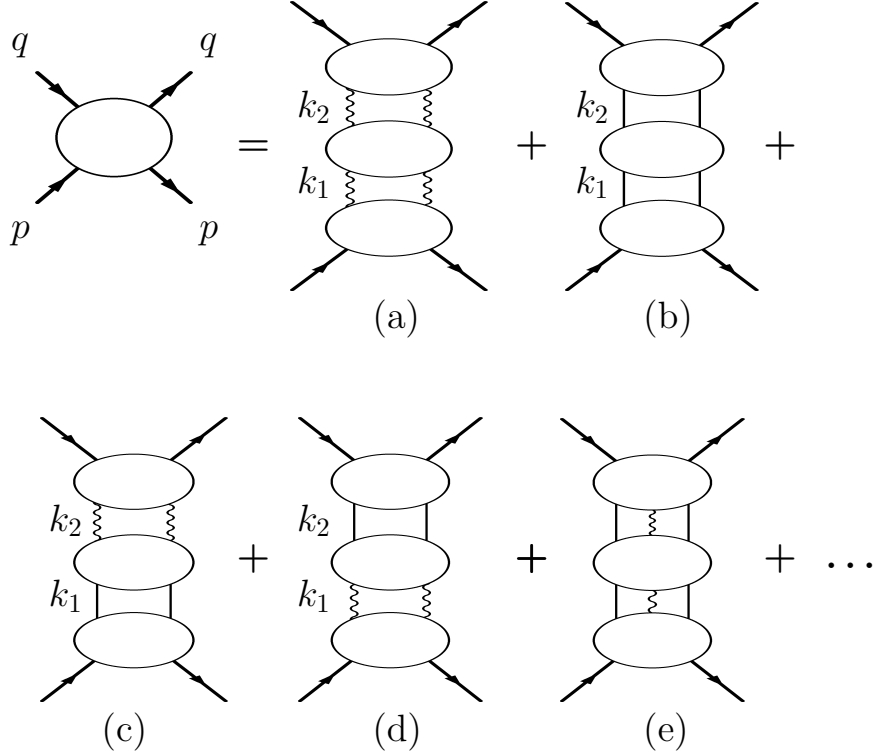} }
\end{picture}
\end{center}
\caption{\label {hfactfig7} Representation of the amplitude $A_h$ through
convolutions of sub-processes.}
\end{figure}

In order to be able to apply Perturbative QCD to this reaction,
should be regarded as consisting of the following two
sub-processes as depicted in
Fig.~\ref{hfactfig7}\\

%%%%%%%%%%%%%%%%%%%%%%%%%%%%%%%%%
%%%%%%%%%%%%%%%%%%%%%%%%%%%%%%%

\textbf{(i)} Each of the colliding hadrons emits one or several partons (the upper and lowest
blobs in Fig.~\ref{hfactfig7}).\\
\textbf{(ii)} High-energy interaction of the emitted
partons (the middle blob in Fig.~\ref{hfactfig7}).

In the present paper we will consider convolutions with two intermediate
partons in the $t$-channel (graphs (a) and (b)) only. This approximation allows us
to write the scattering amplitude $A_h$ of the process of Eq.~(\ref{hscatt}) in
the form of the following convolutions (cf. Eq.~(\ref{factgen})):

\begin{equation}\label{hfactgen}
A_h \approx \sum_{rr'} A_{rr'}
%A_{r,r'}
=  \sum_{rr'} \widetilde{T}^{(r')} (k_2,k'_2,p_2,p'_2) \otimes H^{(r'r)} (k_2,k'_2, k_1,k'_1)
\otimes T^{(r)} (p_1,p'_1, k_1,k'_1) ,
\end{equation}
where the superscripts $r,r'$ refer to the kind of
intermediate two-parton states: two-quark or two-gluon states.
$T^{(r)}$ and $\widetilde{T}^{(r')}$ stand for the lowest and
upper blobs respectively while $H^{(r'r)}$ denotes the middle blob
in Fig.~\ref{hfactfig7}; $k'_{1,2} = q + k_{1,2}$, with $q= p'_1 -
p_1$.

Similarly to the Compton scattering amplitudes considered in Sect.~II, the amplitudes $A_h$ for
hadron scattering can have
either vacuum or non-vacuum quantum numbers in the $t$-channel. In the present paper we focus on the
singlet amplitudes with the vacuum quantum numbers. Then, in what follows we consider
in more detail the amplitude $A_{gg}$ which
%where $r=r'$ and both of them stand for the two-gluon state.
corresponds to graph (a) in
Fig.~\ref{hfactfig7}:

\begin{equation}\label{aggconv}
A_{gg} = \int \frac{d^4 k_1}{(2 \pi)^4} \frac{d^4 k_2}{(2 \pi)^4}
\widetilde{T}^{(g)}_{\mu\mu'}(p_2,p'_2, k_2,k'2)  \frac{1}{k^2_2 k'^2_2}
H^{(gg)}_{\mu\mu'\lambda\lambda'} (k_2,k_1)
\frac{1}{k^2_1 k'^2_1} T^{(g)}_{\lambda\lambda'} (p_1,p'_1,k_1,k'_1),
\end{equation}
where the subscripts $\mu,\mu',\lambda,\lambda'$ denote the gluon polarizations. In order to
account for the leading contributions, we parameterize $T^{(g)}_{\lambda\lambda'}$
and $\widetilde{T}^{(g)}_{\mu\mu'}$ as follows:

\begin{equation}\label{t}
\widetilde{T}^{(g)}_{\mu\mu'} = \frac{2 p_{2 \mu} p_{2 \mu'}}{s}
\widetilde{T},\qquad T^{(g)}_{\lambda\lambda'} =
\frac{2p_{1\lambda} p_{1 \lambda'}}{s} T,\qquad H = \frac{2 p_{2
\mu} p_{2 \mu'}}{s} H^{(gg)}_{\mu\mu'\lambda\lambda'}
\frac{2p_{1\lambda} p_{1 \lambda'}}{s},
\end{equation}
$\widetilde{T},~T,~H$ being scalar functions. We have
dropped the superscripts  here. Using these notations and the
Sudakov parametrization of Eq.~(\ref{sud})  for $k_{1,2}$, we can rewrite
Eq.~(\ref{aggconv}) as follows:

\begin{equation}\label{aggsud}
A_{gg} = \int d \alpha_{1,2} d \beta_{1,2} d k^2_{1,2 \perp}
\widetilde{T}(s\beta_2, k_2^2,k'^2_2)  \frac{1}{k^2_b}
\frac{s'}{k^2} M
\frac{1}{k'^2_a} T (s\alpha_1, k^2_1,k'^2_1),
\end{equation}
with $k^2_r = - s\alpha_r\beta_r - k^2_{r \perp}$.
Let us explain the notations we have introduced in Eq.~(\ref{aggsud}).
We have denoted by $s'$ the invariant energy for the
sub-process of  gluon $2 \to 2$ -scattering: $s' = (k_1 -
k_2)^2 = (k'_1 - k'_2)^2 \approx s \alpha_2 \beta_1$. Throughout
the paper we consider the kinematics where
\begin{equation}\label{regge}
s' \gg  k^2_1, k^2_2, k'^2_1,k'^2_2.
\end{equation}
Let us denote by $k_c$ the momentum with the largest
virtuality: $k^2_c =$max$[k^2_1, k^2_2, k'^2_1,k'^2_2]$.
Then we denote by $k^2_b $ any of
$k^2_2,k'^2_2$, providing $b \neq c$. Similarly,
$k^2_a $ is any of $k^2_1,k'^2_1$, providing $a \neq c$. Finally,  $k^2 =
$min$[k^2_1, k^2_2, k'^2_1,k'^2_2]$. The amplitudes
$T,~\widetilde{T}$ can include both perturbative and
non-perturbative contributions, as can amplitude $M$.
The amplitude $M$ replaces the amplitude $H$
when the momenta $k_a, k_b, k$ are used instead of $k_1, k_2$.
 The integration over $k_{1,2}$
in Eq.~(\ref{aggsud})  runs over the whole phase space, so the
integration region includes the IR-singularities where any of $k^2_1,~k'^2_1$
or $k^2_2,~k'^2_2$ can be equal to zero. Obviously, $k^2_2$ and $k'^2_2$ (and
$k^2_1$ and $k'^2_1$) can be small at the same time only when $q =
0$, i.e.when $t=0$ and therefore the forward kinematics becomes collinear. This
kinematics is the most IR-singular. At $t = 0$
Eq.~(\ref{aggsud}) looks as follows:

\begin{equation}\label{aggcol}
A_{gg} = \int d \alpha_{1,2} d \beta_{1,2} d k^2_{1,2 \perp}
\widetilde{T}(s\beta_2, k^2_2)  \frac{1}{k^2_2}
\left(\frac{s'}{k^2}\right) M
\frac{1}{k'^2_1} T (s\alpha_1,k^2_1)
\end{equation}
where $k^2 =$ min$[k^2_2, k^2_1]$.

The perturbative infrared-sensitive contributions in the amplitude $M$ are logarithms:
$M = M \left(\ln (s'/k^2_1), \ln (s'/k^2_2) \right)$.

\begin{figure}[h]
\begin{center}
\begin{picture}(140,140)
\put(0,0){\epsfbox{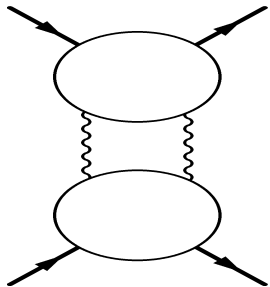} }
\end{picture}
\end{center}
\caption{\label {hfactfig8} Convolution for $A_h$ in the Born approximation.}
\end{figure}

On the other hand, the hadron amplitude $A_{gg}$ must be free of IR problems by definition.
In order to keep Eq.~(\ref{aggcol}) IR-stable, Amplitudes $T$ and $\widetilde{T}$ should
compensate the IR-divergent terms (cf. Eq.~(\ref{tsnsk})):

\begin{equation}\label{tsk}
T \sim (k^2_1)^{1 + \gamma},\qquad \widetilde{T} \sim (k^2_2)^{1 +
\gamma},
\end{equation}
with $\gamma > 0$. When $t \neq 0$, the conditions for the IR stability are weaker
than in Eq.~(\ref{tsk}):

\begin{equation}\label{tnsk}
T \sim (k^2_a)^{\gamma},\qquad \widetilde{T} \sim
(k^2_b)^{\gamma}.
\end{equation}
With the integrations in Eqs.~(\ref{aggsud},\ref{aggcol})
IR-stable and the total energy of the gluon scattering high, the
amplitude $M$ becomes completely perturbative. When $M$ is calculated in the
Born approximation, the amplitude $A_h$ is represented in
Fig.~\ref{hfactfig8}. Let us notice that the Born approximation
means that the blobs in Fig.~\ref{hfactfig8} include
non-perturbative contributions only.

%%%%%%%%%%%%%%%%%%%%%%%%%%%%%%%%%
%%%%%%%%%%%%%%%%%%%%%%%%%%%%%%%

\begin{figure}[h]
\begin{center}
\begin{picture}(260,200)
\put(0,0){\epsfbox{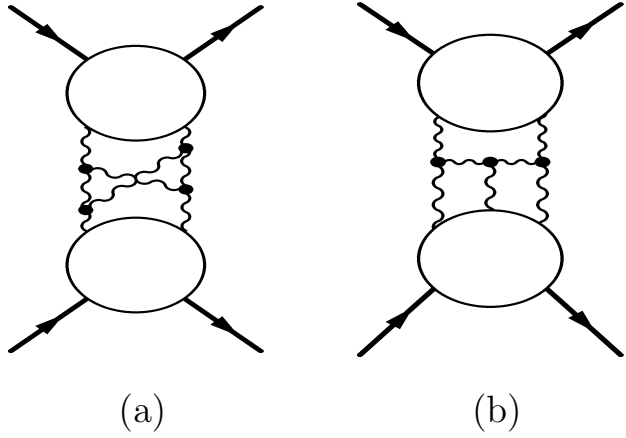} }
\end{picture}
\end{center}
\caption{\label {hfactfig9} Convolutions for $A_h$ beyond the Born approximation.}
\end{figure}

Considering $A_{gg}$ beyond the Born approximation corresponds to adding
the quark and gluon propagators to the Born graph. In doing so,
we account for those graphs which do not involve the upper and lower
blobs, like graph (a) in Fig.~\ref{hfactfig9} but avoid including
the graphs involving the blobs like graph (b).

%%%%%%%%%%%%%%%%%%%%%%%%%%%%%%%%%
%%%%%%%%%%%%%%%%%%%%%%%%%%%%%%%

Such a procedure keeps the blobs totally non-perturbative. The
gauge invariance of the set of retained graphs is proved in
Appendix~B. As a result, we arrive back at graph (a) depicted in
Fig.~\ref{hfactfig7} where, however, the middle blob is
perturbative while the upper and lowest blobs are altogether
non-perturbative. Therefore, the convolutions in
Eqs.~(\ref{aggconv},\ref{aggsud}) complemented by the IR-regulators
of Eqs.~(\ref{tsk}, \ref{tnsk}) represent the hadron scattering amplitude
$A_h$ in terms of the convolutions of
Eqs.~(\ref{aggsud},\ref{aggcol}), with the convolutions being in
the perfect agreement with the concept of the QCD factorization.
Applying the same analysis to the amplitudes $A_{qq}, A_{qg},
A_{gq}$, we arrive to the representation of the hadron amplitude
$A_h$ in terms of the QCD factorization. Obviously, this
factorization differs from the $k_T$- and collinear factorization,
so we again refer to it as the Basic Factorization.

\subsection{Transition to the $k_T$ -factorization}

The transition to  $k_T$ -factorization means performing the
integrations  over $\alpha_1$ and $\beta_2$ in
Eqs.~(\ref{aggsud},\ref{aggcol}) without dealing with the amplitude
$(s'/k^2)M$. This is possible only if these integrations are
restricted as follows:

\begin{equation}\label{abk12}
s \alpha_1\beta_1 \ll k^2_{1 \perp},~~s \alpha_2\beta_2 \ll k^2_{2 \perp}.
\end{equation}
The meaning  of these restrictions is that the virtualities of the
intermediate partons are space-like. This agrees with the known fact
that the virtualities of the ladder partons in the Leading
Logarithmic Approximation are space-like.

Restriction  (\ref{abk12}) allows us to integrate over $\beta_2$
and $\alpha_1$ in Eqs.~(\ref{aggsud},\ref{aggcol}), arriving at
the following expression for $A_h (s,t)$ in $k_T$
-factorization:

\begin{equation}\label{ahktfact}
A_{h} (s,t) = \int d \alpha_2 d \beta_1 d k^2_{1 \perp} d k^2_{2 \perp}
\widetilde{\Psi}(s \alpha_2, k_{2 \perp}^2,k'^2_{2 \perp})  \frac{1}{k^2_{b \perp}}
%\frac{s'}{k^2} M
A^{(pert)}
\frac{1}{k^2_{a \perp}} \Psi (s\beta_1, k^2_{1 \perp},k'^2_{1 \perp}),
\end{equation}
with the  perturbative amplitude $A^{(pert)}$ and non-perturbative
parton distributions $\Psi, \widetilde{\Psi}$ defined as follows:

\begin{equation}\label{am}
A^{(pert)} = \left(s'/k^2_{\perp}\right) M \left(\ln (s'/k^2_{1, \perp}), \ln (s'/k^2_{\perp}),
\ln (s'/k'^2_{1, \perp}), \ln (s'/k'^2_{\perp}) \right),
\end{equation}

\begin{equation}\label{psi}
\widetilde{\Psi} =  \int_{s_1^{(0)}/s}^{k^2_{2 \perp}/s \alpha_2} d \beta_2
\widetilde{T} (s \beta_2, k^2_{2 \perp}, k'^2_{2 \perp}) ,~~ \Psi
=
\int_{s_2^{(0)}/s}^{k^2_{1 \perp}/s \beta_1}
d \alpha_1 T (s \alpha_1, k^2_{1 \perp}, k'^2_{1 \perp})
\end{equation}
with $s_1^{(0)}$ and $s_2^{(0)}$ being the minimal invariant energy of the
upper and lowest blobs, respectively (cf. Eq.~(\ref{stwo})).
Putting $t = 0$, then  taking the imaginary part of
Eq.~(\ref{ahktfact}) and using the Optical theorem, we arrive at
the expression for the total cross-section $\sigma_{tot}$ of
the hadron scattering in  $k_T$ -factorization:
\begin{equation}\label{sigmaktfact}
\sigma_{tot} = \int d \alpha_2 d \beta_1
%\frac{d \alpha_2}{\alpha_2} \frac{d \beta_1}{\beta_1}
\frac{d k^2_{1 \perp}}{k^2_{1 \perp}} \frac{d k^2_{2 \perp}}{k^2_{2 \perp}}
\widetilde{\Phi}(s\alpha_2, k_{2 \perp}^2)
%\frac{1}{k^2_{2 \perp}}
%\frac{s'}{k^2} M
\;\sigma_{tot}^{(pert)} (s', k^2_{1 \perp}, k^2_{2 \perp})
%\frac{1}{k^2_{1 \perp}}
\;\Phi (s\beta_1, k^2_{1 \perp}),
\end{equation}
with the new parton distributions
$\Phi (s\beta_1, k^2_{1 \perp})$ and
$\widetilde{\Phi} (s\alpha_2, k_{2 \perp}^2)$ defined quite similarly to the parton
distributions in Eq.~(\ref{phisnsktfact}):

\begin{equation}\label{psiab12}
\Phi (s\beta_1, k^2_{1 \perp})= \Im \Psi (s\beta_1, k^2_{1
\perp}),\qquad \widetilde{\Phi} (s\alpha_2, k_{2 \perp}^2)= \Im
\tilde{\Psi}(s\alpha_2, k_{2 \perp}^2).
\end{equation}
In Eq.~(\ref{sigmaktfact}) we have used the notation
$\sigma_{tot}^{(pert)}$ for the total cross-section of the
perturbative sub-process describing the scattering of two partons.
$k_T$-factorization was introduced in Ref.~\cite{ktfact} to
replace the collinear factorization in the case when the BFKL
Pomeron\cite{bfkl} is used for the description of  $\sigma_{tot}^{(pert)}$.
Collinear  factorization can be used when alternative methods are
exploited to describe $\sigma_{tot}^{(pert)}$.

\subsection{Transition to  collinear factorization}

The transition from  $k_T$-factorization in
Eq.~(\ref{sigmaktfact}) to  collinear factorization can be done,
assuming a peaked $k_{\perp}$-dependence of the  parton
distributions involved, $\Phi, \widetilde{\Phi}$. This  dependence is
depicted in Fig.~\ref{hfactfig5}. Once we accept this assumption, we can
neglect the $k_{\perp}$-dependence of
$\sigma_{tot}^{(pert)}$ and integrate  only
the parton distributions $\Phi, \widetilde{\Phi}$ over $k_{1,2 \perp}$.  As a result, we arrive at

\begin{equation}\label{sigmacolfact}
\sigma_{tot} = \sum_{k,l} \int d \alpha_2 d \beta_1
%\frac{d \alpha_2}{\alpha_2} \frac{d\beta_1}{\beta_1}
\widetilde{\varphi}_k(s \alpha_2, \mu^2_l)
\;\sigma_{tot}^{(pert)} (s', \mu^2_k)\; \varphi_k (s \beta_1,
\mu^2_k),
\end{equation}
where

\begin{equation}\label{hphimu}
\varphi_k (s \beta_1, \mu^2_k) = \int_{D_k} d k^2_{1 \perp} \Phi (s \beta_1,k^2_{1 \perp}),~~
\widetilde{\varphi}_k (s \alpha_2, \mu^2_k) = \int_{D_k} d k^2_{2 \perp}
\widetilde{\Phi} (s \alpha_2,k^2_{1 \perp})
\end{equation}
where $D_k$ is the region around the maximum of the parton distributions at $k^2_{1,2 \perp} = \mu^2_k$.

The transitions from the basic form of  QCD factorization to  $k_T$ -factorization and to
collinear factorization for the non-singlet component of $A_h$ can be done quite similarly.
Obviously, Eq.~(\ref{sigmacolfact}) differs from the conventional expression

\begin{equation}\label{sigmaconv}
\sigma_{tot} = \int d \alpha_2 d \beta_1
%\frac{d \alpha_2}{\alpha_2} \frac{d\beta_1}{\beta_1}
\widetilde{\phi}_k(s \alpha_2, \mu^2)
\;\sigma_{tot}^{(pert)} (s', \mu^2)\; \phi (s \beta_1, \mu^2),
\end{equation}
where the parton distributions $\phi,~\widetilde{\phi}$ include both perturbative and non-perturbative contributions.

\section{Comparison of Eqs.~(\ref{fsnscolfact},\ref{sigmaktfact}) to the conventional form of
collinear factorization}\label{sect4}

The reduction of the $k_T$ -factorization to collinear factorization led us to
Eq.~(\ref{fsnscolfact}) for the DIS structure functions and Eq.~(\ref{sigmacolfact}) for the total
cross-section. Obviously, these expressions differ a lot from the expressions written in the
conventional form Eq.~(\ref{colfact},\ref{sigmaconv}) of collinear factorization. Let us show how Eqs.~(\ref{fsnscolfact},\ref{sigmacolfact})
can be brought to the conventional form. First of all, notice that
the collinear factorization in Eqs.~(\ref{fsnscolfact},\ref{sigmacolfact}) differs from the conventional
form  in the following
aspects:

\textbf{(A)} There is only one factorization scale $\mu$ in Eqs.~(\ref{colfact},\ref{sigmaconv}) and
and this scale does not bear any
physical meaning, so $\mu$ can be chosen arbitrary. On the contrary,
Eqs.~(\ref{fsnscolfact},\ref{sigmacolfact})
admit the possibility of having several intrinsic scales $\mu_r$,
each corresponds to the maximum $k^2_{\perp} = \mu^2_r$ of the unintegrated parton
distributions.

\textbf{(B)} The parton distributions $\varphi,~\widetilde{\varphi}$ in Eqs.~(\ref{fsnscolfact},\ref{sigmacolfact})
include the
non-perturbative distributions only whereas
the conventional form operates with the distributions $\phi,~\widetilde{\phi}$ where
there are both perturbative and non-perturbative
contributions.

Despite such a considerable difference between these two approaches, it is easy to show that
in fact Eqs.~(\ref{fsnscolfact},\ref{sigmacolfact}) can be brought to the form of Eqs.~(\ref{colfact},\ref{sigmaconv}).
This can be done identically for Eq.~(\ref{fsnscolfact}) and Eq.~(\ref{sigmacolfact}), so in what follows we
will focus on considering Eq.~(\ref{fsnscolfact}).
Obviously, Eq.~(\ref{colfact}) can be written as (cf. Eq.~(\ref{factgen}))

\begin{equation}\label{colfactsymb}
f (x,Q^2)= E (Q^2/\mu^2) \otimes \tilde{\phi} (\mu^2)
\end{equation}
where $E (Q^2/\mu^2)$ denotes any of the appropriate evolution operator for
% of any of available perturbative
%evolution
performing the evolution from $\mu^2$ to $Q^2$. We have dropped its $x$-dependence as unessential.
In particular, it can be the DGLAP -evolution operator\cite{dglap}. Using such notation,
Eq.~(\ref{fsnscolfact}) can be rewritten as follows:

\begin{equation}\label{colfactmkm}
f (x,Q^2)= \sum_k E(Q^2/\mu^2_k) \otimes \phi_k (\mu^2_k) =
E (Q^2/\mu^2) \otimes \sum_k E (\mu^2/\mu^2_k) \otimes \phi_k (\mu^2_k) =
E (Q^2/\mu^2) \otimes \phi (\mu^2),
\end{equation}
with
\begin{equation}\label{phimkm}
\phi (\mu^2) = \sum_k E (\mu^2/\mu^2_k) \otimes \varphi_k (\mu^2_k).
\end{equation}
Eq.~(\ref{colfactmkm}) makes sense if $\mu > \mu_k$. This perfectly agrees with
the actual situation:
the scale $\mu$ in the standard approach
is usually chosen $\sim 1$GeV whereas the intrinsic scales $\mu_k$ are supposedly non-perturbative and
 they are therefore expected to be not far from $\Lambda_{QCD}$.
Eq.~(\ref{colfactmkm}) demonstrates that the conventional approach and our approaches
can be converted each into other when the factorization scale $\mu$ is chosen at the GeV range,
so they are equivalent and indistinguishable when $\mu$ is kept within the GeV range or higher.
In this case the parton distributions should explicitly include perturbative
terms in addition to the non-perturbative ones.
They become distinguishable at smaller values of the factorization scale, close
to $\Lambda_{QCD}$ (where the perturbative component vanishes),
providing there are several (more than one) intrinsic scales $\mu_k$, otherwise they are
again indistinguishable.
However, the difference between them can be found in the framework of the $k_T$ -factorization
where the existence of the maximums in
dependence $\Phi = \Phi (k^2_{\perp})$ can be checked with analysis of experimental data.

\section{Ultraviolet behavior of the parton distributions $T_S,~T_{NS}$}\label{sect5}

The basic factorization convolutions for the Compton amplitudes
(i.e. Eqs.~(\ref{aq},\ref{aqsud}) complemented by Eqs.~(\ref{tsnsk})) and for the hadron amplitudes $A_h$
(i.e. Eqs.~(\ref{aggsud},\ref{aggcol}) complemented by Eqs.~(\ref{tsk})) are
IR-stable. In addition, they should be UV-stable. The UV-stability of the Compton amplitudes was
discussed in detail in Ref.~\cite{egtfact}. It was proved that the convolutions are UV-stable when
the parton distributions $T_S (\alpha),T_{NS}(\alpha)$ at large $|\alpha|$ behave as follows:

\begin{equation}\label{tsnsuv}
T_{NS} \sim |\alpha|^{-1 - h},\qquad T_{S} \sim |\alpha|^{- h},
\end{equation}
with $h > 0$.
%It was also shown that after the transition to the $k_T$-factorization has been done,
%the parton distributions $\Psi_S,~\Psi_{NS}$ should satisfy (\ref{tsnsuv}) as well.
It is easy to show
that the singlet parton distributions $T(\alpha_1),~\widetilde{T} (\beta_2)$
in the convolutions of Eqs.~(\ref{aggsud},\ref{aggcol}) should exhibit absolutely the same UV behavior:

\begin{equation}\label{tsab}
T(\alpha_1) \sim |\alpha_1|^{-h},\qquad \widetilde{T} \sim
\beta_2^{-h}.
\end{equation}

\section{Restrictions on the parton distributions}\label{sect6}

We obtain below the restriction on the parton distributions in the collinear and
$k_T$-factorizations that follow from the integrability of the basic convolutions. We exploit here
the obvious mathematical requirement of stability of the convolutions in both the IR and UV -regions.
These restrictions can also be used as suggestions when PDF fits are constructed.

\subsection{Parton distributions in  $k_T$ -factorization}

The fits for the parton distributions in $k_T$-factorization are commonly fixed from phenomenological
considerations (see Ref.~\cite{fitskt}). Below we give some restrictions on the fits following from
theoretical grounds.
The parton distributions $\Phi$ in $k_T$-factorization are defined in Eq.~(\ref{phisnsktfact})
(see also Eq.~(\ref{psiab12}))
as the integrals of the totally unintegrated parton densities.  Combining this equation with the
restriction of Eq.~(\ref{tsnsk}) (see also Eq.~(\ref{tsk})) on their IR-behavior  leads us to the most general
form of $\Phi_{NS},~\Phi_S$:

\begin{eqnarray}\label{phisnsktgen}
\Phi_{NS} &=& (k^2_{\perp})^{\gamma} \chi^{(1)}_{NS} (\beta, k^2_{\perp})
+ (k^2_{\perp})^{\gamma - h} \beta^{h}\chi^{(2)}_{NS}(\beta, k^2_{\perp}),
\\ \nonumber
\Phi_S &=& (k^2_{\perp})^{1+ \gamma} \chi^{(1)}_{S} (\beta, k^2_{\perp})
+ (k^2_{\perp})^{2 + \gamma - h} \beta^{-1 + h} \chi^{(2)}_{S}(\beta, k^2_{\perp}),
\end{eqnarray}
with the requirement

\begin{equation}\label{gammah}
\gamma - h > 0
\end{equation}

We remind that the positive parameters $\gamma$ and $h$ were introduced in Eqs.~(\ref{tsnsk},\ref{tsnsuv})
for the Compton amplitudes
to guarantee their
IR and UV stability, respectively. When the Compton amplitudes are considered in basic
factorization, $\gamma$ and  $h$ are independent. However, they proved to be related by Eq.~(\ref{gammah})
when basic factorization has been reduced to $k_T$ -factorization: Eq.~(\ref{gammah}) ensures the IR-stability
of the structure functions in $k_T$-factorization.
Eq.~(\ref{phisnsktgen}) shows that the parton densities in $k_T$-factorization
consist of two terms, each involves different power factors of $k^2_{\perp}$ and the last $k^2_{\perp}$-factor involves the longitudinal variable $\beta$.
In addition, they involve arbitrary functions  $\chi^{(1,2)}_{S,NS}$. All $\chi^{(1,2)}_{S,NS} \to$ const at $k^2_{\perp} \to 0$.
%becomes independent of $k^2_{\perp}$.
$\Phi_{S,NS}$ originate from $T_{S,NS}$, so the requirements of the UV stability in Eqs.~(\ref{tsnsuv},\ref{tsab})
mean that the fits for $\Phi_{NS}$ should not contain the factors $\beta^{-a}$, with $a > 0$, whereas the
requirement on fits for
$\Phi_S$ is less severe: they
can contain such factors, providing $a<1$. However, sometimes the same parton distributions contribute to
the singlet and non-singlet constructions. For example, this is true for the flavor singlet components of the structure functions $F_1$
(addressed as the singlet in the present paper)
and $F_2$ (addressed as the non-singlet): they both involve the same parton distributions.
In such cases, the singlet fits should obey the more severe requirement for the non-singlets.
In order to make possible the transitions to the collinear factorization, the $k_{\perp}$-dependence of $\chi^{(1,2)}_{S,NS}$
should be of a peaked (for example, Gaussian) form.

\subsection{Parton distributions in collinear factorization}

The conventional fits\cite{fitscolfact} for the parton distributions in collinear factorization are also constructed from purely
phenomenological considerations. Below we impose some restrictions on these fits arising from
integrability the factorization convolutions.
The standard fits for the initial parton densities $\delta q, \delta g$ in  collinear factorization
are known to
include
a normalization $N$, the singular factors $x^{-a}$, with $a > 0$,
and  regular terms. For example,

\begin{equation}\label{fit}
\delta q
%, \delta g =
= N x^{-a} (1 - x)^b (1 + c x^d)\,,
\end{equation}
where the parameters $N, a, b, c, d$ are positive. They are specified from analysis of experimental data.
Although such expressions do not look explicitly like the ones
obtained with the perturbative methods, nonetheless there are two options to study: \\
\textbf{(i)} The singular factors $x^{-a}$ appear as a result of total resummations of the perturbative
contributions.
In this case we identify the fits with the parton distributions $\phi$ containing both perturbative and non-perturbative
contributions. The factors $x^{-a}$ in such distributions mimic the resummation of either the leading logarithms  (see \cite{egtg1})
or sub-leading logarithms (see \cite{efr}). In the both cases the singular factors can be
removed from the fits when the resummation is accounted for  because the resummation of
logarithms of $x$ can be absorbed by the coefficient functions. \\
\textbf{(ii)}  The whole fits Eq.~(\ref{fit}), including the singular factors $x^{-a}$, have the non-perturbative origin.
In this case we
identify the fits in Eq.~(\ref{fit}) with the non-perturbative distributions $\varphi$ defined in Eqs.~(\ref{phisnscolfact},\ref{hphimu})
and apply to them the restrictions of Eqs.~(\ref{tsnsuv},\ref{tsab}). These restrictions
exclude the use of the singular factors in the expressions for the non-singlet
structure functions $F_2, F^{NS}_1, g_1$, etc
and also suppress the singular factors with $a > 1$ in the
expressions for the singlet $F_1$. However, the parton  distributions used for
$F_1$ and $F_2$ are identical,
therefore the suppression of the singular factors with $a > 0$ can be
applied to all structure functions, including the singlet $F_1$. The singular
factors $x^{-a}$ in the DGLAP fits for initial parton densities should
be removed from the fits because they contradict to the
integrability of the basic convolutions of the Compton amplitudes. Let us remark that the removal of the factors $x^{-a}$
from the fits and replacing such factors by the total resummation of the logarithms can reduce the fits in Eq.~(\ref{fit}) down to
much simpler expressions like $N (1 - x)^b$ or, when the starting values of $x$ are small, down to constants.

\section{Summary}\label{sect7}

We have shown that the scattering amplitudes of the processes Eq.~(\ref{comptscatt}) and
Eq.~(\ref{hscatt}) can be represented as basic convolutions where the perturbative and non-perturbative
contributions are separated (i.e. located in different blobs). The basic convolutions correspond to
concept of  QCD factorization but they are more general than collinear and $k_T$-factorization.
The perturbative components of the basic convolutions are off-shell and therefore they are not
gauge-invariant. However, their gauge invariance is restored at high energy (small $x$).
%It is well-known that at high energy the transverse momenta of the ladder partons care greater
When the virtualities of the intermediate partons are space-like, the basic factorization can be
approximately reduced to  $k_T$-factorization. In its turn, $k_T$-factorization can approximately
be reduced to collinear factorization if the peaked dependence  on $k_{\perp}$ (see Fig.~5) is assumed for
the unintegrated parton distributions and the parton distributions are altogether non-perturbative.
The sharper the peaks in Fig.~5 are, the higher is the accuracy of the reduction.
We stress that this hypothesis can be checked by analysis of experimental data.
In contrast to the conventional scenario of  collinear factorization,
%where the factorization scale can be
%chosen arbitrarily and parton distributions include both perturbative and non-perturbative contributions,
we have arrived at the more involved form of collinear
factorization, with one of several (intrinsic) scales $\mu_r$ corresponding
to the maxima of the unintegrated parton distributions.  Using perturbative evolution
to increase the scale, we have brought
this complicated many-scale picture to the conventional form, where only one arbitrary scale
is used explicitly and the parton distributions include both perturbative and non-perturbative contributions.
The integrations in the basic convolutions
run over the whole phase space but they must yield finite results.
Exploiting this obvious requirement allows us to  obtain restrictions on the parton distributions
in both collinear and $k_T$-factorization. We have shown that the parton distributions in
$k_T$-factorization should consist of two terms, with different powers of $k^2_{\perp}$.
Then, we exclude the use of factors $x^{-a}$  from the fits
for the parton distributions both in the collinear and $k_T$-factorization. Finally, let us remark that
the results obtained in the present paper can easily be extended to inelastic hadron scattering at high energy.

\acknowledgements{ We are grateful to B.~Webber for interesting
discussions. We are also grateful to S.~Alekhin and G.~Lykasov for useful correspondence.
The work is partly supported by Grant RAS 9C237,
Russian State Grant for Scientific School RSGSS-65751.2010.2 and
EU Marie-Curie Research Training Network under contract
MRTN-CT-2006-035505 (HEPTOOLS).}

\appendix
\section{Gauge invariance of the Compton scattering amplitude}

\subsection{Notation for Compton scattering}

Let us consider the Compton scattering

\begin{equation}\label{compt}
\gamma (q) + q (p) \to \gamma (q') + q (p') .
\end{equation}
We denote by $A_{\mu\nu}$ the scattering amplitude of the process (\ref{compt}).
Gauge invariance states that

\begin{equation}\label{qaugeinvgen}
q_{\mu}A_{\mu\nu} = q_{\nu}A_{\mu\nu} = 0.
\end{equation}

\subsection{Compton scattering amplitude in the Born approximation}

In the Born approximation the scattering amplitude $A_{\mu\nu}$ is
represented by the graphs depicted in Fig.~\ref{hfactfig3}.

\begin{equation}\label{aborn}
A_{\mu\nu} = \frac{<p'| \gamma_{\nu} (\hat{p}+ \hat{q} + m)\gamma_{\mu}|p>}{(p + q)^2 - m^2}
+ \frac{<p'| \gamma_{\mu} (\hat{p}'- \hat{q} +m)\gamma_{\nu}|p>}{(p'- q)^2 - m^2}
\end{equation}

\subsection{Gauge invariance of the Compton amplitude in arbitrary kinematics}

It follows from Eq.~(\ref{aborn}) that 

\begin{eqnarray}\label{aborninv}
q_{\mu}A_{\mu\nu} = -  \frac{<p'| \gamma_{\nu}\hat{q}(\hat{p}- m)|p>}{(p + q)^2 - m^2}
-  \frac{<p'|(\hat{p}'- m)\hat{q}\gamma_{\nu}|p>}{(p'- q)^2 - m^2}
\\ \nonumber
+ <p'|\gamma_{\nu} |p> \left[\frac{q^2 + 2pq}{(p + q)^2 - m^2}~
- \frac{q^2 - 2p'q}{(p' - q)^2 -m^2}\right]
\end{eqnarray}

Eq.~(\ref{aborninv}) can be re-written as follows:

\begin{eqnarray}\label{aborninv1}
q_{\mu}A_{\mu\nu} = -  \frac{<p'| \gamma_{\nu}\hat{q}(\hat{p} - m)|p>}{(p + q)^2 - m^2}
-  \frac{<p'|(\hat{p}' - m)\hat{q}\gamma_{\nu}|p>}{(p'- q)^2 - m^2}
\\ \nonumber
- <p'|\gamma_{\nu} |p> \left[\frac{p^2 -m^2}{(p + q)^2 - m^2}~
- \frac{p'^2 - m^2}{(p' - q)^2 -m^2}\right]
\end{eqnarray}

Obviously, Eq.~(\ref{aborninv1})  yields a zero result only when
the incoming and outgoing quarks are on-shell. In contrast, the
photons can be either on-shell or off-shell.

%\section{Gauge invariance for the forward Compton amplitude}

Let us focus on considering forward Compton scattering.

\subsection{Forward Compton amplitude in the Born approximation}

In this case

\begin{equation}\label{abornf}
A_{\mu\nu} = \frac{<p| \gamma_{\nu} (\hat{p}+ \hat{q} + m)\gamma_{\mu}|p>}{(p + q)^2 - m^2}
+ \frac{<p| \gamma_{\mu} (\hat{p}- \hat{q} +m)\gamma_{\nu}|p>}{(p- q)^2 - m^2}.
\end{equation}
Obviously, the forward Compton amplitude is gauge-invariant when the quark is on-shell.
However, now we allow it to be off-shell. In order to study
gauge invariance for an off-shell quark,
we rewrite $A_{\mu\nu} $ as follows:

\begin{equation}\label{a123}
A_{\mu\nu}= A_{\mu\nu}^{(1)}+ A_{\mu\nu}^{(2)} + A_{\mu\nu}^{(3)}
\end{equation}
where

\begin{eqnarray}\label{abornf1}
A_{\mu\nu}^{(1)} = \left[\frac{<p| \gamma_{\nu} \hat{p} \gamma_{\mu}|p>}{(p + q)^2 - m^2}
+ \frac{<p| \gamma_{\mu} \hat{p}\gamma_{\nu}|p>}{(p- q)^2 - m^2}\right],
\\ \nonumber
A_{\mu\nu}^{(2)} = \left[\frac{<p| \gamma_{\nu} \hat{q}\gamma_{\mu}|p>}{(p + q)^2 - m^2}
- \frac{<p| \gamma_{\mu} \hat{q}\gamma_{\nu}|p>}{(p- q)^2 - m^2} \right],
\\ \nonumber
A_{\mu\nu}^{(3)} =  m \left[\frac{<p| \gamma_{\nu} \gamma_{\mu}|p>}{(p + q)^2 - m^2}
+ \frac{<p| \gamma_{\mu} \gamma_{\nu}|p>}{(p- q)^2 - m^2}\right]
\end{eqnarray}

Let us multiply Eq.~(\ref{a123}) by $q_{\nu}$ and represent the result as follows:

\begin{equation}\label{g123}
G_{\mu} \equiv q_{\nu}A_{\mu\nu}= G_{\mu}^{(1)}+ G_{\mu}^{(2)} + G_{\mu}^{(3)},
\end{equation}
with

\begin{equation}\label{g1}
G_{\mu}^{(1)} \equiv q_{\nu} A_{\mu\nu}^{(1)} =
\left[\frac{<p| \hat{q} \hat{p} \gamma_{\mu}|p>}{(p + q)^2 - m^2}
+ \frac{<p| \gamma_{\mu} \hat{p}\hat{q}|p>}{(p- q)^2 - m^2} \right],
\end{equation}

\begin{equation}\label{g2}
G_{\mu}^{(2)} \equiv q_{\nu} A_{\mu\nu}^{(2)} =
q^2 <p|\gamma_{\mu}|p>
\left[\frac{1}{(p + q)^2 - m^2}
- \frac{1}{(p- q)^2 - m^2} \right],
\end{equation}

\begin{equation}\label{g3}
G_{\mu}^{(3)} \equiv q_{\nu} A_{\mu\nu}^{(3)} = m
\left[\frac{<p|\hat{q}\gamma_{\mu}|p>}{(p + q)^2 - m^2}
+ \frac{<p|\gamma_{\mu} \hat{q}|p>}{(p- q)^2 - m^2} \right].
\end{equation}

Now let us simplify the Dirac structures in (\ref{g1}):
\begin{eqnarray}\label{dirac}
\gamma_{\mu}\hat{q}\hat{p} &=& -\imath \epsilon_{\mu\lambda\rho \sigma} q_{\lambda} p_{\rho}
\gamma_5 \gamma_{\sigma} + q_{\mu} \hat{p} - p_{\mu} \hat{q}  + pq \gamma_{\mu}
\\ \nonumber
\hat{q} \hat{p} \gamma_{\mu} &=& \imath \epsilon_{\mu\lambda\rho \sigma} q_{\lambda} p_{\rho}
\gamma_5 \gamma_{\sigma} + q_{\mu} \hat{p} - p_{\mu}\hat{q} + pq \gamma_{\mu}
\end{eqnarray}

This makes it possible to represent  $G_{\mu}^{(1)}$ as a sum of
symmetrical and antisymmetrical contributions:

\begin{equation}\label{g1sas}
G_{\mu}^{(1)} = G_{\mu}^{(1A) + }G_{\mu}^{(1S)},
\end{equation}
with

\begin{equation}\label{g1a}
G_{\mu}^{(1A)} = \imath \epsilon_{\mu\lambda\rho \sigma} q_{\lambda} p_{\rho}
\frac{<p|\gamma_5 \gamma_{\sigma}|p>}{w}
\left[\frac{1}{1 -x -z} + \frac{1}{1 +x +z} \right]
,
\end{equation}
where $x = - q^2/w,~z= m^2/w$ and $w=2pq$.

Similarly,

\begin{equation}\label{g1s}
G_{\mu}^{(1S)} =
\frac{[q_{\mu} <p| \hat{p}|p> - p_{\mu}<p|\hat{q}|p> + pq <p|\gamma_{\mu}|p>]}{w}
\left[\frac{1}{1-x -z} - \frac{1}{1+x + z} \right]
.
\end{equation}

In order to retain the CP-invariance in (\ref{g1a}), the term $<p|\gamma_5 \gamma_{\sigma}|p> $
must yield the spin contribution $\sim S_{\sigma}$. This converts the antisymmetrical
factor in   (\ref{g1a}) into

\begin{equation}\label{epsilon1}
\epsilon_{\mu\lambda\rho \sigma} q_{\lambda} p_{\rho} S_{\sigma}.
\end{equation}

In order to simplify (\ref{g1s}) we use the approximation

\begin{equation}\label{pp}
<p|\gamma_{\mu}|p> \approx p_{\mu},
\end{equation}
which allows us to re-write (\ref{g1s}) as

\begin{equation}\label{g1s1}
G_{\mu}^{(1S)} \approx
[q_{\mu} p^2 - p_{\mu}pq + p_{\mu} pq]\frac{<p|p>}{w}
\left[\frac{1}{1-x -z} - \frac{1}{1+x + z} \right].
\end{equation}

Obviously,  the expressions on the r.h.s. of Eqs.~(\ref{g1a},
\ref{g1s1}) are not equal to zero, which means that
 gauge invariance for $A_{\mu \nu}^{(1)}$ is broken. The same is true for $A_{\mu \nu}^{(2)}$
 and $A_{\mu \nu}^{(3)}$ when the quark is off-shell.

\subsection{Restoration of the gauge invariance for forward Compton scattering at small $x$ }

In the high-energy limit where $w \gg |q^2|, |(p^2-m^2)|$, i.e. where both
\begin{equation}\label{smallx}
x \ll 1
\end{equation}
and
\begin{equation}\label{smallz}
z \ll 1,
\end{equation}
the expression in the square brackets in Eq.~(\ref{g1s1}) becomes small and therefore in
this kinematics
 $G_{\mu}^{(1S)} \sim \max[x,z] \approx 0$. Then, in this kinematic region the spin $S_{\rho}$ becomes
mostly longitudinal, so the expression in (\ref{epsilon1}) is
nearly zero and therefore $G_{\mu}^{(1A)} \approx 0$.
Eq.~(\ref{g2}) demonstrates  that $G_{\mu}^{(2)}$ in region
(\ref{smallx}) is proportional to $x$, so $G_{\mu}^{(2)} \approx
0$.  Finally, let us notice that $G_{\mu}^{(3)} \sim m$, so it
becomes small in the small-$x$ region (\ref{smallx}).
Therefore, the gauge invariance of the forward Compton amplitude,
although broken when the off-shell quark is off-shell, is restored in
the limit of small $x$ and $z$ for the unpolarized Compton
scattering amplitudes and for those spin-dependent amplitudes,
in which the spin is longitudinal.

\subsection{Accounting for radiative corrections}

The most important radiative corrections in the small-$x$ region
are logarithmic ones. The arguments of the logarithms can be
chosen as $w/p^2$ and $Q^2/p^2$. Accounting  for the radiative
corrections with logarithmic accuracy does not destroy the
Born structure of the forward Compton amplitude.
%and
Then, it is obvious that
$\ln (w/p^2)$ and $\ln (Q^2/p^2)$ can be large only if $p^2 \ll Q^2, w$. So, the main contributions
in the small-$x$ region come from quark virtualities obeying (\ref{smallz}).
Therefore the total resummation of such contributions does not break the
gauge invariance of the Compton amplitude at small $x$.
This gives us the right to consider the
factorization convolutions with the two-parton
intermediate states only and to neglect all other intermediate states with greater
numbers of partons without breaking gauge invariance at small $x$.

\section{Gauge invariance for quark-quark scattering in collinear kinematics}

\begin{figure}[h]
\begin{center}
\begin{picture}(300,120)
\put(0,0){\epsfbox{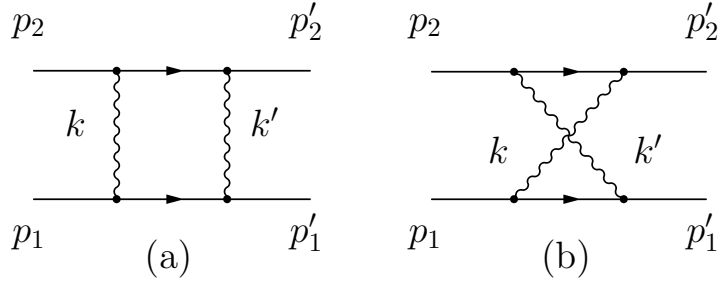} }
\end{picture}
\end{center}
\caption{\label {hfactfig10} Colorless
contribution to the quark scattering in the lowest order.}
\end{figure}

Let us consider the quark-quark scattering

\begin{equation}\label{qq}
q(p_1) + q (p_2) \to q (p'_1) + q (p'_2)
\end{equation}
in collinear kinematics where $p_1 \approx p'_1$, $p_2 \approx
p'_2$. In what follows we will focus on the colorless, in the
$t$-channel, part of the scattering amplitude of this process. In
other words, we will consider the amplitude with vacuum
quantum numbers in the $t$-channel. In the lowest-order
approximation the colorless part  of the scattering amplitude $A_0$
is depicted in Fig.~\ref{hfactfig10}.

%%%%%%%%%%%%%%%%%%%%%%%%%%%%%%%%%
%%%%%%%%%%%%%%%%%%%%%%%%%%%%%%%

We write it as follows:

\begin{equation}\label{aus}
A_0 = A_s + A_u,
\end{equation}
where

\begin{equation}\label{as}
A_s \sim \frac{<p_1|\gamma_{\nu} (\hat{p}_1 - \hat{k} + m) \gamma_{\mu}|p_1>
<p_2|\gamma_{\nu'} (\hat{p}_2 + \hat{k} + m) \gamma_{\mu'}|p_2>}
{\left((p_1 - k)^2 - m^2\right) \left((p_2 + k)^2 - m^2\right)}
\frac{d_{\mu'\mu} (k)}{k^2} \frac{d_{\nu'\nu} (k)}{k^2}
\end{equation}
and

\begin{equation}\label{au}
A_u \sim \frac{<p_1|\gamma_{\nu} (\hat{p}_1 - \hat{k} + m) \gamma_{\mu}|p_1>
<p_2|\gamma_{\mu'} (\hat{p}_2 + \hat{k} + m) \gamma_{\nu'}|p_2>}
{\left((p_1 - k)^2 - m^2\right) \left((p_2 - k)^2 - m^2\right)}
\frac{d_{\mu'\mu} (k)}{k^2} \frac{d_{\nu'\nu} (k)}{k^2}
\end{equation}
As our object is to study the gauge invariance of $A_0$, we have
omitted in Eqs.~(\ref{as},\ref{au}) all factors unessential for
that. As  is well-known, the leading contributions to $A_s$ and
$A_u$ come from the terms $\sim p_1, p_2$ in the numerators of
(\ref{as}, \ref{au}), with $k$ and $m$ neglected. The remaining
structures are obviously symmetrical under the replacements $\mu
\rightleftharpoons\nu$ and $\mu' \rightleftharpoons\nu'$.
Therefore we can write them as

\begin{equation}\label{aslead}
A_s^L =
\frac{S_{\mu\nu} (p_1)
%\frac{<p_1|\gamma_{\nu} \hat{p}_1\gamma_{\mu}|p_1>
%<p_2|\gamma_{\nu'} \hat{p}_2 \gamma_{\mu'}|p_2>}
S_{\mu'\nu'} (p_2)}
{\left(k^2 - 2p_1k +(p^2_1 - m^2)\right) \left(k^2 + 2p_2k +(p^2_2 -m^2)\right)}
\frac{d_{\mu'\mu} (k)}{k^2} \frac{d_{\nu'\nu} (k)}{k^2}
\end{equation}

and

\begin{equation}\label{aulead}
A_u^L =
\frac{S_{\mu\nu} (p_1)S_{\mu'\nu'} (p_2)}
{\left((k^2 - 2p_1k + (p^2_1 -m^2)\right) \left(k^2 - 2p_2 k +(p^2_2-m^2)\right)}
\frac{d_{\mu'\mu} (k)}{k^2} \frac{d_{\nu'\nu} (k)}{k^2},
\end{equation}
with

\begin{equation}\label{s}
S_{\mu\nu} (p_1) = <p_1|\gamma_{\nu} \hat{p}_1 \gamma_{\mu}|p_1>,~
S_{\mu'\nu'} (p_1) = <p_2|\gamma_{\mu'} \hat{p}_2 \gamma_{\nu'}|p_2> .
\end{equation}

Gauge invariance means that the replacement of
$d_{\nu'\nu}(k)$ by $k_{\nu} \phi_{\nu'} (k)$,  $\phi$ being
an arbitrary function, yields  a zero result. Multiplying
$S_{\mu\nu} (p_1) $ by $k_{\nu}$ and substituting the result in
Eqs.~(\ref{aslead},\ref{aulead}) proves the breaking of gauge
invariance. However, in the high-energy limit where

\begin{equation}\label{inv}
|2p_1k|, |2p_2k| \gg |k^2|, |(p^2-m^2)|
\end{equation}
gauge invariance is restored.
This is the appropriate region to produce the leading logarithmic contribution. This means that
accounting for these contributions does not destroy the gauge invariance of the amplitude $A_0$.

\end{document}